\documentclass[aps,prx,twocolumn,groupedaddress]{revtex4-2}
\usepackage{mathtools}
\usepackage[colorlinks]{hyperref}
\newcommand{\ket}[1]{|#1\rangle}
\newcommand{\bra}[1]{\langle #1 |}

\begin{document}

\title{Self-Ordering of Individual Photons in Waveguide QED and Rydberg-Atom Arrays}

\author{Ole A. Iversen}
\author{Thomas Pohl}
\affiliation{Center for Complex Quantum Systems, Department of Physics and Astronomy, Aarhus University, Ny Munkegade 120, DK-8000 Aarhus C, Denmark}

\date{\today}

\begin{abstract}
The scattering between light and individual saturable quantum emitters can induce strong optical nonlinearities and correlations between individual light quanta. Typically, this leads to an effective attraction that can generate exotic bound states of photons, which form quantum mechanical precursors of optical solitons, as found in many optical media. Here, we study the propagation of light through an optical waveguide that is chirally coupled to three-level quantum emitters. We show that the additional laser-coupling to a third emitter state not only permits to control the properties of the bound state but can even eliminate it entirely. This makes it possible to turn an otherwise focussing nonlinearity into a repulsive photon-photon interaction. We demonstrate this emerging photon-photon repulsion by analysing the quantum dynamics of multiple photons in large emitter arrays and reveal a dynamical fragmentation of incident uncorrelated light fields and self-ordering into regular trains of single photons. These striking effects expand the rich physics of waveguide quantum electrodynamics into the domain of repulsive photons and establish a conceptually simple platform to explore optical self-organization phenomena at the quantum level. We discuss implementations of this setting in cold-atom experiments and propose a new approach based on arrays of mesoscopic Rydberg-atom ensembles. 
\end{abstract}

\maketitle

\section{Introduction}
Light-matter interactions give rise to a plethora of nonlinear phenomena \cite{boyd03} that can originate from intensity-dependent refraction and absorption of the optical medium. Well-known examples include the self-focussing of laser beams \cite{Kelley65}, the formation optical solitons \cite{Hasegawa89} or modulational instabilities that cause filamentation of light \cite{Peccianti03,Tai86}. These effects follow from the attractive nature of optical nonlinearities that occur in most materials, and they often require high light intensities, as the nonlinear response is typically rather small.

On the other hand, substantial research efforts are currently being directed towards reaching nonlinearities that can operate at the lowest possible intensities down to the level of individual photons \cite{quantum-non-linear-review}. The corresponding effective photon-photon interactions can find applications in optical quantum technologies such as quantum information processing \cite{Ralph15,Tiarks19,Murray17} and quantum communication \cite{Borregaard20}, and they offer exciting prospects for exploring quantum many-body physics with light \cite{Noh16}. A number of different experiments have successfully demonstrated the generation and application of quantum optical nonlinearities, e.g. based on optical resonators \cite{Birnbaum05,Hacker16} or nanoscale waveguides \cite{chang18-review,turschmann19-review,sheremet21-review} to enhance the interaction between light and individual emitters or by utilizing long-range interactions in ensembles of atomic quantum emitters \cite{firstenberg2016-review,murray2016-review}. The resulting photon-photon interactions in these experiments are usually attractive, which has been used to demonstrate the emergence of photonic bound states in an extended nonlinear medium \cite{firstenberg13,liang18} or from the coupling to single quantum emitters \cite{javadi15,hallet18,Jeannic21,paris-mandoki17,Stiesdal18}. Apart from few recent exceptions \cite{cantu20,clark20}, repulsive photon-photon interactions have, however, remained scarce and difficult to realize.

\begin{figure}[t!]
\begin{center}
	\includegraphics[width=\columnwidth]{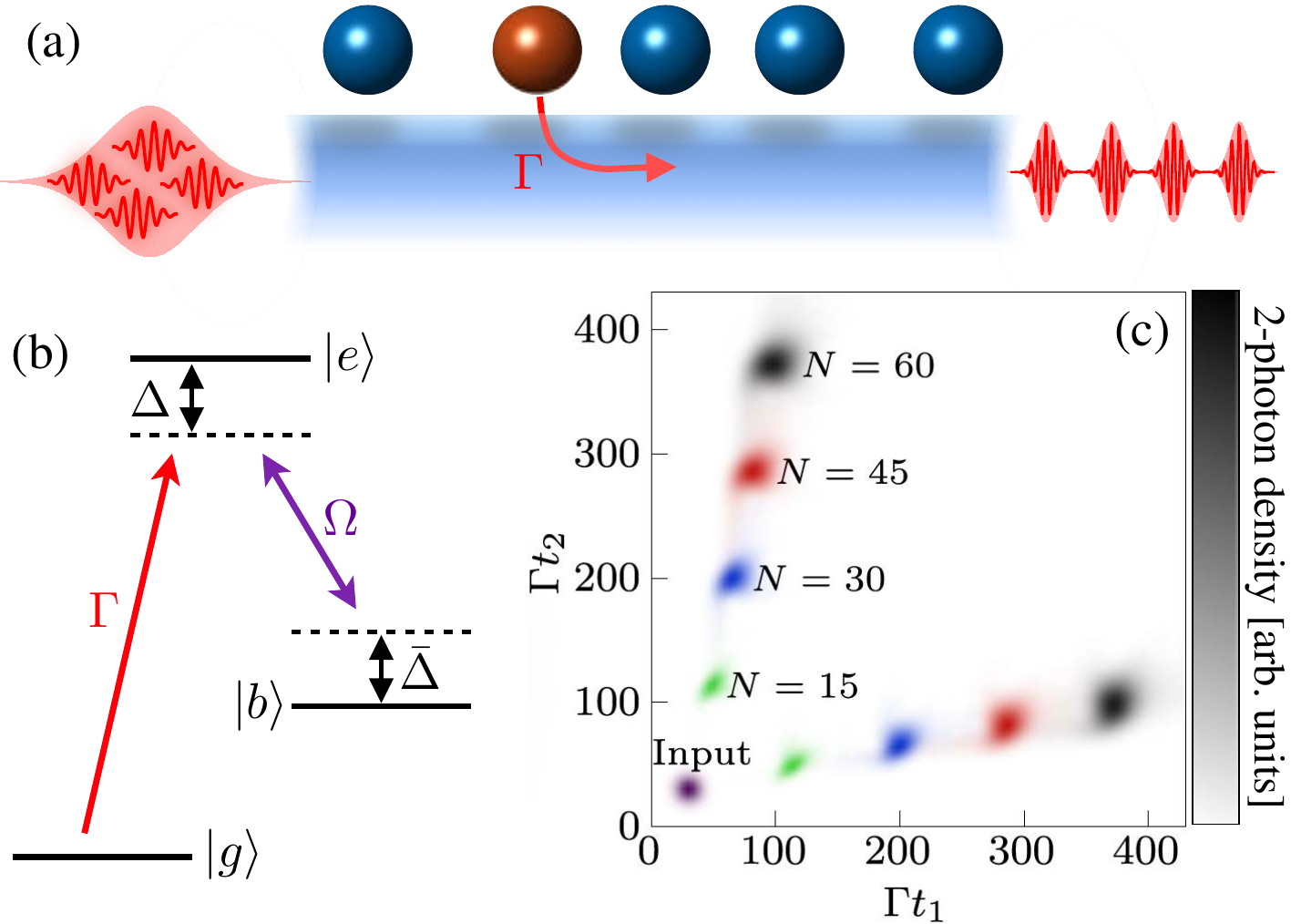}
	\caption{(a) The propagation of light through a chiral chain of three-level emitters can turn an incident pulse of uncorrelated photons into a correlated state of individual photons. Panel (b) depicts the underlying level scheme for chiral photon emission into a single waveguide mode with a rate $\Gamma$ and additional classical control-field driving with a Rabi frequency $\Omega$. The frequency detunings $\Delta$ and $\bar\Delta$ for the respective transitions are also indicated. The calculated two-photon density is shown in panel (c) for different indicated propagation lengths, i.e. numbers of atoms $N$, in the chain for $\Omega = 0.25\Gamma$, $\Delta = 0.5\Gamma$, and $\bar{\Delta} = 0.2\Gamma$. The results show how an incident two-photon product state dynamically fragments into a train of spatially separated single-photon pulses, detected at distinct times $t_1$ and $t_2$.}
	\label{fig:system}
\end{center}
\end{figure}

In this work, we demonstrate the emergence of strong repulsive interactions between multiple photons in a conceptually simple system of three-level emitters that are chirally coupled to an optical waveguide. Unidirectional coupling \cite{chiral-qed-review} between photons and two-level emitters has been achieved with nanoscale waveguides coupled to cold atoms \cite{Vetsch10,petersen14,mitsch14,bechler18} or semiconductor quantum dots \cite{Lodahl15,sollner15,Coles16,JalaliMehrabad20,dewhurst10} as well as with mesoscopic Rydberg-atom ensembles that can realize effective saturable two-level emitters coupled to a single free-space photonic mode \cite{paris-mandoki17}. The propagation of light in chains of such two-level emitters is attracting growing theoretical interest \cite{pletyukhov12,stannigel12,ringel14,ramos14,pichler15,eldredge16,kornovan17,mirza17,mahmoodian18,jen19_driven,jen20_subradiance,Jones20,jen20_single_excitation,Zhong20,mahmoodian20,kumlin20,poudyal20,Wang20,Zhang20,Poshakinskiy21,poshakinskiy2021,fang15,kumlin18,mirza16,buonaiuto19,needham19,jen20_phase-diagram,zhang2021universal,asenjo-carcia17}, and the generation of nonclassical light \cite{prasad20} as well as the emergence of two-photon bound states \cite{Stiesdal18,Jeannic21} has been demonstrated in recent experiments. Here, we show that the introduction of an additional control-field coupling (see Fig. \ref{fig:system}) has profound consequences for the nonlinear behaviour of a system with chiral coupling. In fact, it turns out that adding a third state of the quantum emitter can fundamentally change its  scattering physics and remove the two-photon bound state, which otherwise emerges as a generic scattering eigenstate of two-level systems \cite{shen07-prl,shen07-pra} and forms a quantum mechanical precursor of optical solitons \cite{mahmoodian20}. Correspondingly, the attractive photon interaction familiar from chains of two-level emitters is turned into an effective photon repulsion, which can cause fragmentation of light into individual quanta [see Fig. \ref{fig:system}(c)] and induce spontaneous self-ordering phenomena at the level of single photons.

The article is organized as follows. After introducing the basic physics of the system in Section \ref{sec:model}, we discuss the scattering between two photons and a single three-level emitter in Section \ref{sec:eigenstates} and analyse the properties of the two-photon bound state and their consequences for scattering off multiple emitters. Section \ref{sec:input-output} provides a brief description of the numerical approach to simulate the quantum dynamics of more than two photons. Results of such simulations are presented in Sections \ref{sec:two-pulse} and \ref{sec:multi-pulse}, where we discuss the buildup of strong photon-photon correlations from initially uncorrelated multi-photon light pulses. An experimental implementation via collective photon coupling to cold atomic ensembles is proposed and analysed in Section \ref{sec:superatoms}, while a brief conlusion and outlook on our work is given in Section \ref{sec:conclusion}.

\section{Chiral chains of three-level emitters} \label{sec:model}
Figure \ref{fig:system}(a) illustrates the considered system of $N$ quantum emitters that are coupled to a one-dimensional single-mode waveguide. A single propagating photonic mode generates a strong optical coupling between the ground state $\ket{g}$ and an excited state $\ket{e}$ of the emitters with an emission rate $\Gamma$ and no loss to other photonic modes [see Fig. \ref{fig:system}(b)]. The excited state is further coupled to a long-lived state $\ket{b}$ by a classical control field with Rabi frequency $\Omega$. 

The total Hamiltonian $\hat{H} = \hat{H}_{\rm ph} + \hat{H}_{\rm em} + \hat{H}_{\rm int}$, which describes this system, is split into a photonic Hamiltonian $\hat{H}_{\rm ph}$, a part describing the emitters $\hat{H}_{\rm em}$, and the light-matter interaction $\hat{H}_{\rm int}$. We consider a chiral emitter-photon coupling such that there is only a single  mode propagating in one direction, as described by 
\begin{equation}
	\hat{H}_{\rm ph} = -i\int dx\,\hat{a}^{\dagger}(x)\partial_x\hat{a}(x), \label{eq:Hph}
\end{equation}
where $\hat{a}^{\dagger}(x)$ creates a photon at position $x$, and where we have set the speed of light in the waveguide to unity and $\hbar = 1$, such that space and time have the same units. The emitter Hamiltonian is given by
\begin{equation}
	\hat{H}_{\rm em} = \sum_{j=1}^{N}\left[\omega_e \hat{\sigma}_{ee}^{(j)} + (\omega_b + \bar{\omega})\hat{\sigma}_{bb}^{(j)} + \Omega\left(\hat{\sigma}_{be}^{(j)} + \hat{\sigma}_{eb}^{(j)}\right)\right], \label{eq:Hem}
\end{equation}
where $\sigma_{\alpha\beta}^{(j)} \equiv \ket{\alpha_j}\bra{\beta_j}$, while $\omega_e$ and $\omega_b$ denote the energies of the excited states $\ket{e}$ and $\ket{b}$, respectively.
The photons couple to the emitters according to 
\begin{equation}
	\hat{H}_{\rm int} = \sum_{j=1}^N\left[\int dx\, \sqrt{\Gamma}\delta(x-x_j)\hat{a}(x)\hat{\sigma}_{eg}^{(j)} + {\rm H.c.}\right]. \label{eq:Hint}
\end{equation}
The incident photon frequency $\omega$ and the frequency $\bar{\omega}$ of the coupling field are parametrized by the single-photon detuning $\Delta = \omega_e - \omega$ and two-photon detuning $\bar{\Delta} = \omega_b + \bar{\omega} - \omega$, which are indicated in Fig. \ref{fig:system}(a).

\section{Two-photon transport and the disappearance of the photonic bound state} \label{sec:eigenstates}
The propagation of two photons through a chain of quantum emitters can be solved by finding the eigenstates of the scattering matrix ($S$-matrix) for a single emitter \cite{mahmoodian18}. The $S$-matrix yields the outgoing state $\ket{{\rm out}} = \hat{S}\ket{{\rm in}}$ in terms of the incident two-photon state $\ket{{\rm in}}$, and its eigenstates are invariant under repeated scattering off multiple emitters. The local nature of the emitter-photon interaction suggests that one of these states is localized with respect to distance between the two incident photons, and can be considered a photon-bound state  \cite{Xu13}, as the eigenstates of the $S$-matrix remain unchanged under repeated interactions as the photons propagate through the chain. Such photon bound states generically emerge from the scattering off two-level emitters \cite{shen07-prl,shen07-pra}, while a recent analysis showed that three-level emitters can give rise to yet another type of two-photon state that contains both localized and delocalized contributions \cite{iversen21}, similar to metastable bound states or scattering resonances in atomic collisions \cite{negative-ion-review,fano-nanoscale-review}. 

\begin{figure}[t!]
\begin{center}
	\includegraphics[width=\columnwidth]{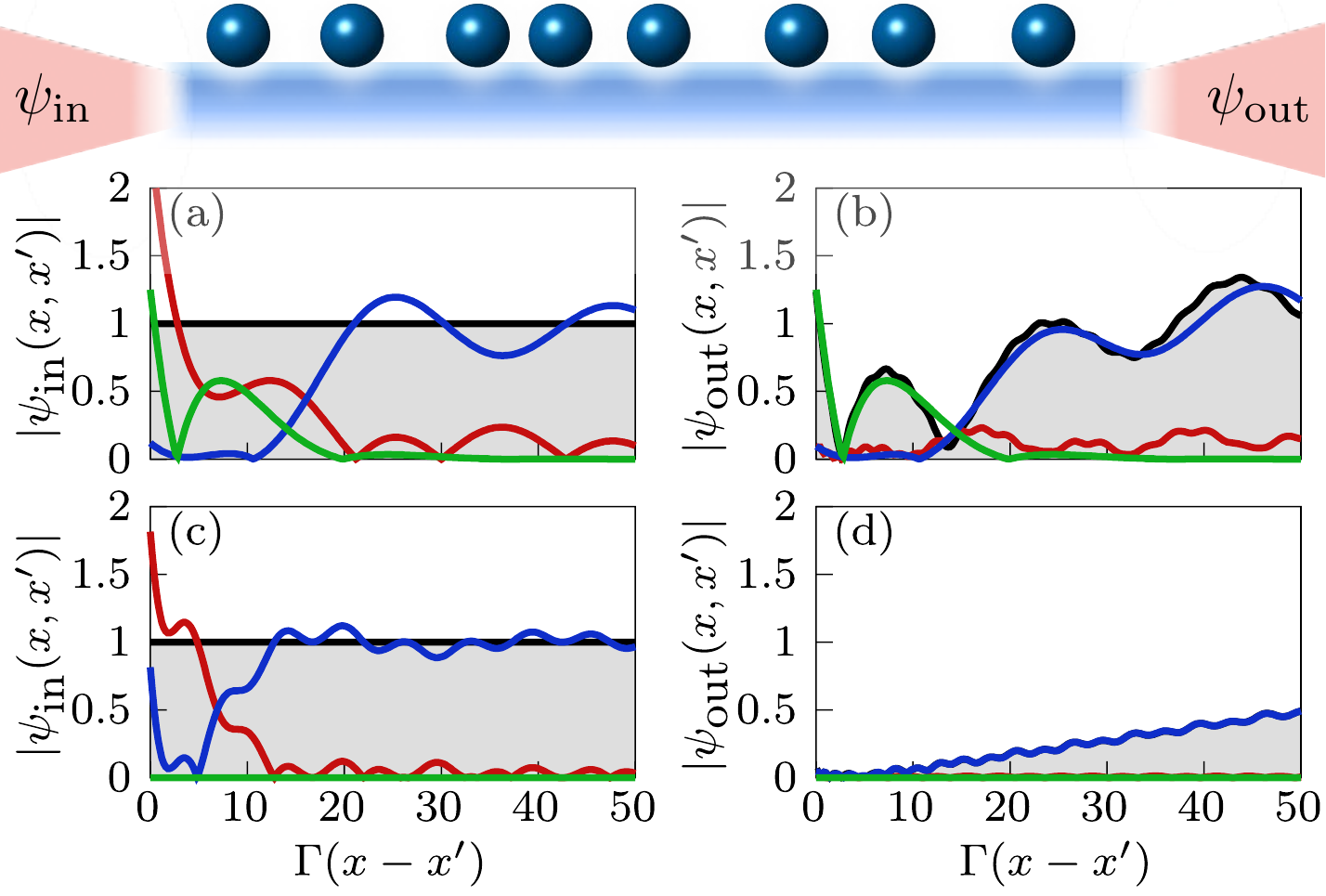}
	\caption{Propagation of a two-photon Fock state through a chain of $N = 100$ emitters. The incident state [panels (a) and (c)] is expanded in scattering states (blue lines), scattering resonance states (red lines), and a bound state (green lines). These types of $S$-matrix eigenstates combine to the full state, indicated by the black lines and shaded area. In panel (a), a bound state exists and contributes to the decomposition of the full state, whereas no bound state is found for the parameters in panel (c). The outgoing states and their expansions, corresponding to panels (a) and (c), after propagation through the chain are displayed in panels (b) and (d), respectively. Panels (a) and (b) are obtained for $\Omega = 0.52\Gamma$, whereas panel (c) and (d) are obtained for $\Omega = 0.35\Gamma$. The detunings are $\Delta = 0.5\Gamma$ and $\bar{\Delta} = 0.4\Gamma$ in all panels.}
	\label{fig:nobound}
\end{center}
\end{figure}

As we will discuss below, it turns out that the additional control-field coupling to a third state of the emitters can even cause the bound state to disappear entirely with profound consequences for the transport properties of a chain of such emitters. 
The general eigenstate for two photons at positions $x$ and $x'$ can be written
\begin{equation}
	\Psi_{E,\nu}(x,x') = e^{iEr_c}F_{E,\nu}(r),
\end{equation}
where $r_c = (x+x')/2$ is the center of mass coordinate of the photons, $r = x - x'$ is their distance, $E$ denotes the total energy, and $\nu$ parametrizes the relative momentum of the two photons. The structure of the underlying two-photon Schr\"odinger equation dictates the general form \cite{iversen21}
\begin{equation}
	F_{E,\nu}(r) = A_{E,\nu}e^{-i\nu r} + B_{E,\nu}e^{i\nu r} + C_{E,\nu} e^{-i\tilde{\nu}r} + D_{E,\nu}e^{i\tilde{\nu}r}, \label{eq:F}
\end{equation}
for the relative two-photon eigenstate, where the value of $\tilde{\nu}$ is determined by the two quantum numbers $E$ and $\nu$. Equation \eqref{eq:F} provides the eigenstates for $r < 0$, while the part for $r > 0$ is obtained by bosonic symmetry. Three distinct classes of eigenstates are afforded by Eq. \eqref{eq:F}.
\begin{itemize}
\item[$(i)$] If $\nu$ and $\tilde{\nu}$ are real, the resulting superposition of plane waves yields a delocalized scattering state.
\item[$(ii)$] If $\nu$ and $\tilde{\nu}$ are both complex and $B_{E,\nu} = D_{E,\nu} = 0$, Eq. \eqref{eq:F} yields an exponentially localized solution corresponding to a two-photon bound state. 
\item[$(iii)$] If $\nu$ is real, $\tilde{\nu}$ is complex, and $D_{E,\nu} = 0$ one obtains localized and delocalized contributions that form the resonance-like state discussed above.
\end{itemize}
These different types of $S$-matrix eigenstates have a profound effect on the nonlinear transport properties of the chain, as revealed by the two-photon amplitude of the transmitted light (see Fig. \ref{fig:nobound}). By definition, the state $|{\rm out}\rangle=\hat{S}|{\rm in}\rangle=\int dE \mathrlap{\sum}\int\limits_{\nu} \lambda(E,\nu)\langle E,\nu \vert \rm in\rangle | E,\nu \rangle$ generated by one emitter is fully determined by the eigenstates $| E,\nu \rangle$ and eigenvalues $\lambda(E,\nu)$ of the $S$-matrix and repeated application of this relation yields the wave function of the two-photon output \cite{mahmoodian18}
\begin{equation}
	\psi_{\rm out}(x,x') = \int dE \mathrlap{\sum}\int\limits_{\nu} \lambda(E,\nu)^N\Psi_{E,\nu}(x,x') \langle E,\nu \vert \rm in\rangle \label{eq:decompose}
\end{equation}
from a chain of $N$ emitters in terms of the single-emitter eigenstates of the $S$-matrix. From this expression, we can gain some intuitive insights into the role of the different eigenstates for the two-photon propagation. In fact, the effect of each eigenstate onto the transmittted field $\psi_{\rm out}(x,x')$ is fully determined by the decomposition of the incident two-photon state for which we consider a cw-field with the wave function $\psi_{\rm in}(x,x') = e^{2i\omega r_c}$. Since the $S$-matrix eigenvalues merely contribute a phase rotation with $|\lambda(E,\nu)|=1$, the $\nu$-integration leads to a dephasing of the continuum eigenstates, and effectively diminishes their contribution at small photon distances $|x-x'|$ as the length $N$ of the chain increases. As a result, the bound state with a discrete value of $\nu$ eventually dominates the short-distance behavior. 

Two-level emitters feature a single bound state and a continuum of plane wave states \cite{shen07-pra,shen07-prl} such that long chains generically generate bunched light in the absence of photon losses as $N$ increases \cite{mahmoodian18,iversen21}. The situation changes qualitatively for three-level systems, where the addition of the scattering-resonance states can suppress the bound-state contribution $\langle E,\nu_{\rm b} \vert \rm in\rangle$ to the eigenstate decomposition of the two-photon input. This effect can lead to a transition from a bunched photon output to nonclassical light with substantial antibunching as the length $N$ of the chain increases \cite{iversen21}. 

Surprisingly, a more detailed analysis reveals a broad parameter region in which the photon bound state ceases to exist entirely, i.e. where no solution Eq. \eqref{eq:F} with complex values of $\nu$ and $\tilde{\nu}$ and with $B_{E,\nu} = D_{E,\nu} = 0$ is found. While the interaction with a single emitter still generates localized two-photon output states, consistent with previous findings \cite{roy11,zheng12,roy14} and more general arguments \cite{Xu13}, these localized contributions do not stem from a true photon bound state but rather the two-photon continuum including the scattering resonances. As a result, any localized two-photon input state will eventually dissolve during propagation through a chain of multiple emitters.

Following the above discussion, the elimination of the two-photon bound state should thus enable robust generation of highly antibunched light in a chain of three-level emitters. This is illustrated in Fig. \ref{fig:nobound}, showing the eigenstate decomposition of the input and output state for two cases with different control field amplitudes but otherwise identical parameters. The absence of the bound state for the smaller value of the Rabi frequency $\Omega$ indeed leads to profoundly different photon correlations with near-perfect antibunching of the transmitted light.  

\begin{figure}[t!]
\begin{center}
	\includegraphics[width=\columnwidth]{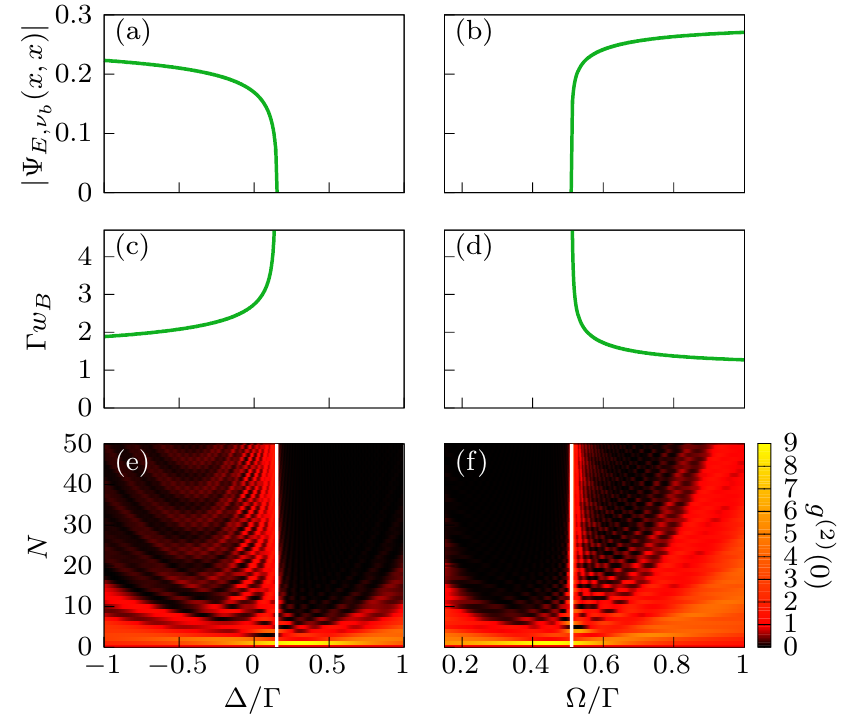}
	\caption{Panels (a)--(d) illustrate the dependence of the two-photon bound state on the single-photon detuning $\Delta$ and the control-field Rabi frequency $\Omega$. The two-photon amplitude $|\Psi_{E,\nu_b}(x,x)|$ at zero photon distance is shown in (a) and (b). It vanishes  at a critical value of $\Delta$ and $\Omega$, indicating the disappearance of the bound state. The root mean square width $w_B$, depicted in (c) and (d), diverges at these values, showing that the bound state indeed broadens and eventually vanishes entirely. Panels (e) and (f) display the profound effects this behavior has on the emerging correlations of photons that propagate through the chain. These panels show the two-photon correlation function at zero delay time for various numbers $N$ of emitters. The white lines mark the disappearance of the bound state found in panels (a)--(d) around which we find strong bunching associated with the bound-state broadening and near-perfect antibunching in the parameter regime with no two-photon bound state. The two-photon detuning is $\bar{\Delta} = 0.4\Gamma$, and panels (a), (c), and (e) are obtained for $\Omega = 0.35\Gamma$, while (b), (d), and (f) correspond to $\Delta = 0.5\Gamma$.}
	\label{fig:transition}
\end{center}
\end{figure}

As shown in Figs. \ref{fig:transition}(a) and (b), the two-photon bound state disappears abruptly as we vary the amplitude of the control field or the single-photon detuning, whereby it broadens as one approaches the critical values $\Omega_c$ and $\Delta_c$ for the Rabi frequency and the single-photon detuning, respectively, as shown in Figs. \ref{fig:transition}(c) and (d). This is concomitant with an abrupt change of the two-photon correlations $g^{(2)}(0)=|\psi_{\rm out}(x,x)|^2/|\psi_{\rm in}(x,x)|^2$ [Figs. \ref{fig:transition}(e) and (f)]. It first rises close to the transition, due to the larger overlap of the broadened bound state with the delocalized input state, and undergoes a sharp transition to $g^{(2)}(0)\approx 0$ as we cross the critical values of the control field parameters. 

The demonstrated ability to optically control and even eliminate the two-photon bound state suggests a viable approach to control the nonlinearity and effective interactions between multiple propagating photons, as we shall discuss below.

\begin{figure}[b!]
\begin{center}
	\includegraphics[width=\columnwidth]{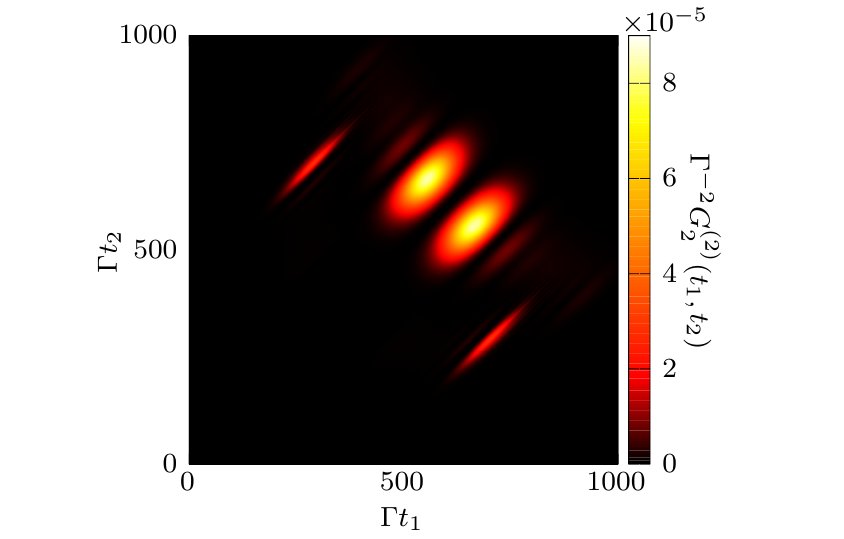}
	\caption{The two-photon density $G_2^{(2)}(t_1,t_2)$ for a long incident two-photon pulse with $\Gamma\sigma = 80$ after propagating through a chain of $N = 60$ emitters. The resulting photon correlations indicate strong antibunching around $t_1 \approx t_2$. The results are obtained for $\Omega = 0.35\Gamma$, $\Delta = 0.5\Gamma$, and $\bar{\Delta} = 0.4\Gamma$.}
	\label{fig:G2-long}
\end{center}
\end{figure}

\section{Simulating multiple photons} \label{sec:input-output}
While the above analysis permits an efficient description of very large numbers of emitters, it becomes exceedingly demanding \cite{rupasov84,yudson85,shen15} for more than two photons. We therefore adapt an alternative approach to the scattering problem that facilitates treatment of coherent incident fields with arbitrary strengths and moderate numbers of emitters \cite{caneva15,manzoni17}. Hereby, the photon dynamics is solved formally within the Markov approximation to obtain an effective Master equation \cite{pichler15}
\begin{equation}
\partial_t\hat{\rho}=-i\left(\hat{H}_{\rm eff}\hat{\rho}-\hat{\rho}\hat{H}_{\rm eff}^{\dagger}\right)+\Gamma\sum_{j,\ell}\hat{\sigma}_{ge}^{(j)}\hat{\rho}\hat{\sigma}_{eg}^{(\ell)} \label{eq:master}
\end{equation}
for the density matrix $\hat{\rho}$ of the quantum emitters from which the transmitted field can be reconstructed according to
\cite{caneva15,manzoni17}
\begin{equation}
	\hat{a}_{\rm out} = \mathcal{E}_{\rm in}(t) - i\sqrt{\Gamma}\sum_{j=1}^{N}\hat{\sigma}_{ge}^{(j)}, \label{eq:input-output-chiral}
\end{equation}
where $\mathcal{E}_{\rm in}(t)$ is the field envelope of the incident coherent light pulse and the creation operator $\hat{a}^\dagger_{\rm out}(t)$ describes the transmitted photons in a frame rotating with the carrier frequency of the input field. The resulting effective Hamiltonian is \cite{stannigel12,pichler15}
\begin{eqnarray}
	\hat{H}_{\rm eff} &=& \hat{H}_{\rm em} + \sqrt{\Gamma} \sum_{j=1}^N \mathcal{E}_{\rm in}(t)\left(\hat{\sigma}_{eg}^{(j)} + \hat{\sigma}_{ge}^{(j)}\right)\nonumber\\
	&& - i\frac{\Gamma}{2}\sum_{j = 1}^N \hat{\sigma}_{ee}^{(j)} - i\Gamma\sum_{j > \ell}\hat{\sigma}_{eg}^{(j)}\hat{\sigma}_{ge}^{(\ell)}, \label{eq:Heff}
\end{eqnarray}
where Eq. \eqref{eq:Hem} is given by 
\begin{equation}
	\hat{H}_{\rm em} = \sum_{j=1}^N \left[\Delta\hat{\sigma}_{ee}^{(j)} + \bar{\Delta}\hat{\sigma}_{bb}^{(j)} + \Omega\left(\hat{\sigma}_{be}^{(j)} + \hat{\sigma}_{eb}^{(j)}\right)\right] \label{eq:Hem-rot}
\end{equation}
in the rotating frame. We solve Eq. \eqref{eq:master} by means of Monte Carlo wave function simulations \cite{molmer96} from which one can reconstruct complete information about the multiphoton light field from Eq. \eqref{eq:input-output-chiral}. 

\begin{figure}[t!]
\begin{center}
	\includegraphics[width=\columnwidth]{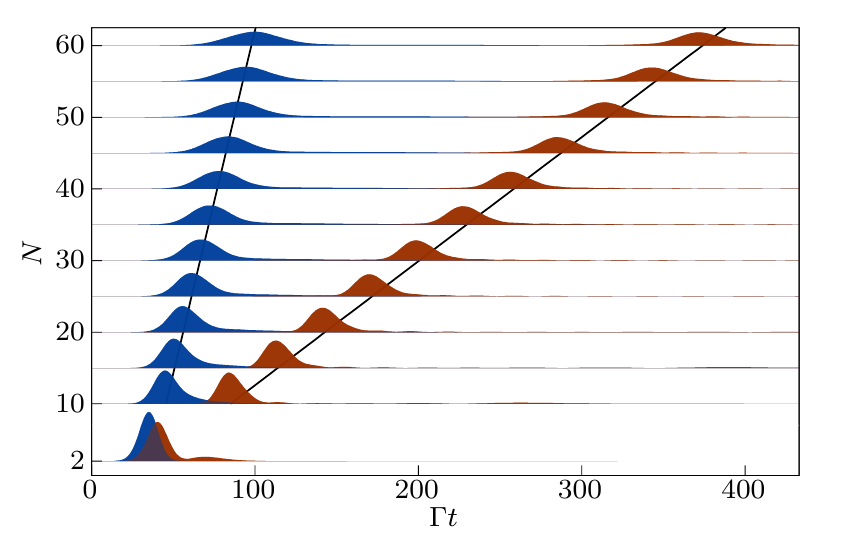}
	\caption{Pulse propagation of the leading and trailing photons as defined in Eqs. \eqref{eq:Pfirst} and \eqref{eq:Psecond}. The strong emerging photon repulsion causes a complete fragmentation into spatially distinct single-photon pulses with a distance that increases linearly with the chain length $N$. The departure times of the generated single-photon pulses are in excellent agreement with analytical results (black lines) based on the different group delay of single photons due to energy-conserving scattering processes in the initial part of the chain (see Appendix \ref{sec:delay}). The depicted data are the same as in Fig. \ref{fig:system}(c) and are obtained for $\Omega = 0.25\Gamma$, $\Delta = 0.5\Gamma$, $\bar{\Delta} = 0.2\Gamma$, and the pulse width $\sigma = 10\Gamma^{-1}$.}
	\label{fig:power-2photon}
\end{center}
\end{figure}

In the absence of losses, one can, moreover, infer the propagation dynamics of incident photon-number states from such simulations. To this end, we simulate the dynamics under very weak coherent driving with an average photon number $\bar{n}\ll1$ and subsequently postselect on a given photon number $k$. For example, this can be accomplished by computing the $k$-photon density 
\begin{align}
	\MoveEqLeft
	G^{(k)}(t_1,\dots,t_k) \nonumber\\
	&= \left\langle \hat{a}_{\rm out}^{\dagger}(t_1)\cdots\hat{a}_{\rm out}^{\dagger}(t_k) \hat{a}_{\rm out}(t_k)\cdots\hat{a}_{\rm out}(t_1)\right\rangle \label{eq:Gn}
\end{align}
from the stochastic wave function simulations \cite{molmer96,caneva15}. Choosing the incident pulse  sufficiently weak, the predominant contribution to the $k$-photon density stems from the $k$-photon component of the coherent photon state. In the absence of photon losses, the computed density thus corresponds to the transmitted state of a $k$-photon Fock-state input. In practice, one can reduce the driving field amplitude $\mathcal{E}_{\rm in}$ until the results converge. In order to obtain the reduced $k$th-order density for an incident Fock state  with a higher photon number $n>k$, one first needs to calculate the $n$th-order density and subsequently trace over the remaining times
\begin{equation}
	G_{n}^{(k)}(t_1,\dots,t_k) = \mathcal{N}_k \int dt_{k+1}\cdots dt_{n}\, G^{(n)}(t_1,\dots,t_{n}), \label{eq:Gn_n'}
\end{equation}
where the pre-factor $\mathcal{N}_k$ ensures that the norm $\int dt_1\dots dt_k G_n^{(k)}(t_1,\dots,t_k)=n$ yields the number of considered photons.

Being primarily interested in the output from an $n$-photon Fock state enables reduction of the Hilbert space dimension for the driven emitters by truncating the basis by omitting all many-body states with more than $n$ excited emitters. This yields a significant reduction from $3^N$ to ${\sim}N^n$ states and thereby facilitates simulations of large systems. The simulations can be further optimized by successively reducing the size of the Hilbert space following each projection in Eq. \eqref{eq:Gn}, which reduces the relevant photon number by $1$ when calculating the $k$-photon density within the Monte Carlo wave function approach. Altogether, this makes it possible to simulate the propagation of up to $4$ photons in comparatively long chains with $N\sim60$ emitters and analyze the resulting nonlinear multiphoton dynamics, as we shall discuss in the following.

\begin{figure}[b!]
\begin{center}
	\includegraphics[width=\columnwidth]{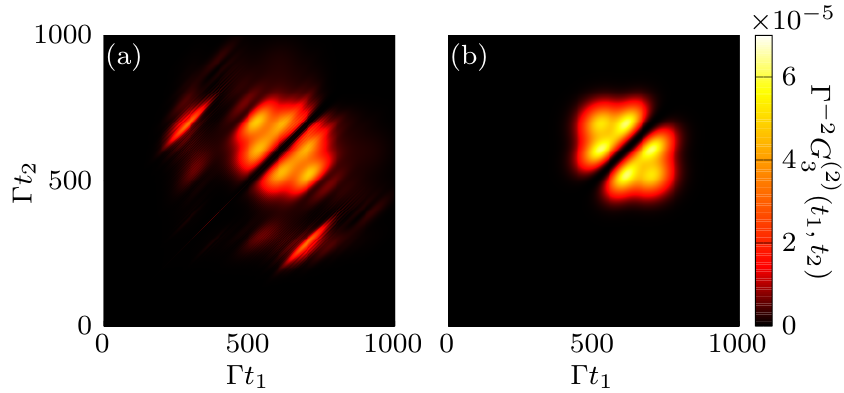}
	\caption{(a) Two-photon density of a three-photon pulse after propagating through a chain of $N = 60$ emitters, with an  incident pulse width of $\Gamma\sigma = 80$. The Rabi coupling is $\Omega = 0.35\Gamma$ and the detunings are $\Delta = 0.5\Gamma$ and $\bar{\Delta} = 0.4\Gamma$. For comparison, panel (b) shows the calculated two-photon density of a correlated reference pulse as given by Eq. \eqref{eq:E-math} with $p = 1.2$, $\Gamma\mu_c = 610$, $\Gamma\sigma_c = 45$, $\Gamma\sigma_{r1} = 100/\sqrt{p}$, and $\sigma_{r2} = 2\sigma_{r1}$. }
	\label{fig:3photon-G2}
\end{center}
\end{figure}

\section{Two-photon fragmentation} \label{sec:two-pulse}
We first consider the dynamics of two photons. To this end, we study the propagation of an incident two-photon pulse with the envelope
\begin{equation}
	\mathcal{E}_{\rm in}(t) = \frac{\sqrt{\bar{n}}}{(\pi\sigma^2)^{1/4}}\exp\left[-\frac{t^2}{2\sigma^2}\right]. \label{eq:Ein}
\end{equation}
Figure \ref{fig:G2-long} shows the transmitted two-photon density through an array of $60$ emitters, which is obtained for a rather long incident pulse with a width of $\Gamma\sigma = 80$. Similar to the above results for cw fields [cf. Fig. \ref{fig:nobound}(d)], one finds pronounced anticorrelations with a vanishing density at small time delays between the two photons.

While this indicates the emergence of repulsive photon interactions, their effects can be revealed far more strikingly for shorter pulses. Here it is important to note that the antibunching feature can extend to very large delays in the output for long pulses. This makes it possible to choose shorter photon pulses whose frequency spectrum lies within the frequency range for which strong antibunching appears, while the pulse duration is still shorter than the extend of the anticorrelation feature. This may suggest that complete fragmentation of a shorter incident pulse into two separate single-photon pulses can be possible for appropriate parameters. Indeed, Fig. \ref{fig:power-2photon} shows a two-photon state fragmenting into a pair of separate single-photon pulses, whose anticorrelation feature, in fact, extends even further than for a corresponding long pulses.

To illustrate this remarkable propagation dynamics, we display in Fig. \ref{fig:power-2photon} the reduced one-photon densities
\begin{align}
	P_{2,1}(t) &= \int_{t}^{\infty} dt_2\, G_{2}^{(2)}(t,t_2), \label{eq:Pfirst}\\
	P_{2,2}(t) &= \int_{-\infty}^{t} dt_2\, G_{2}^{(2)}(t,t_2), \label{eq:Psecond}
\end{align}
which describe the power profiles of the first and second photon in the propagating two-photon pulse, respectively. The photon-photon interaction induced by the first few emitters leads to a significant pulse distortion and causes a complete splitting into spatially separate single-photon pulses after scattering off around $10$ emitters. Eventually, this process generates a train of two individual single-photon pulses that propagate at different constant group velocities through the remainder of the chain with weak overall distortion. The overall two-photon dynamics thus appears equivalent to an effective finite-range interaction that fragments the initial two-body state into repelling one-body wave packets that eventually separate and propagate at constant velocities. 

This simplified picture of a relative momentum kick generated by the effective photon repulsion can be further corroborated by a spectral analysis of the transmitted light (see Appendix \ref{sec:delay}). This shows that the two interacting photons undergo an energy conserving scattering process into different frequencies ${\sim}\omega\pm0.5\Gamma$, where the single-photon pulse distortion is weak, as evindent from Fig. \ref{fig:power-2photon}. As a result of the frequency shifts, the photons acquire different group velocities (see Appendix \ref{sec:delay})
\begin{equation}
	v_g = \frac{(\Omega^2 - \Delta\bar{\Delta})^2 + \Gamma^2 \bar{\Delta}^2/4}{\Gamma(\Omega^2 +\bar{\Delta}^2)}, \label{eq:vg}
\end{equation}
due to the corresponding shifts $\Delta\mp0.5\Gamma$ and $\bar{\Delta}\mp0.5\Gamma$ of the detunings. This simple estimate of the propagation speed and the underlying picture of a two-photon collision offer a remarkably accurate account for the observed overall photon dynamics, as shown by the solid black lines in Fig. \ref{fig:power-2photon}. 

\begin{figure*}
\begin{center}
	\includegraphics[width=\textwidth]{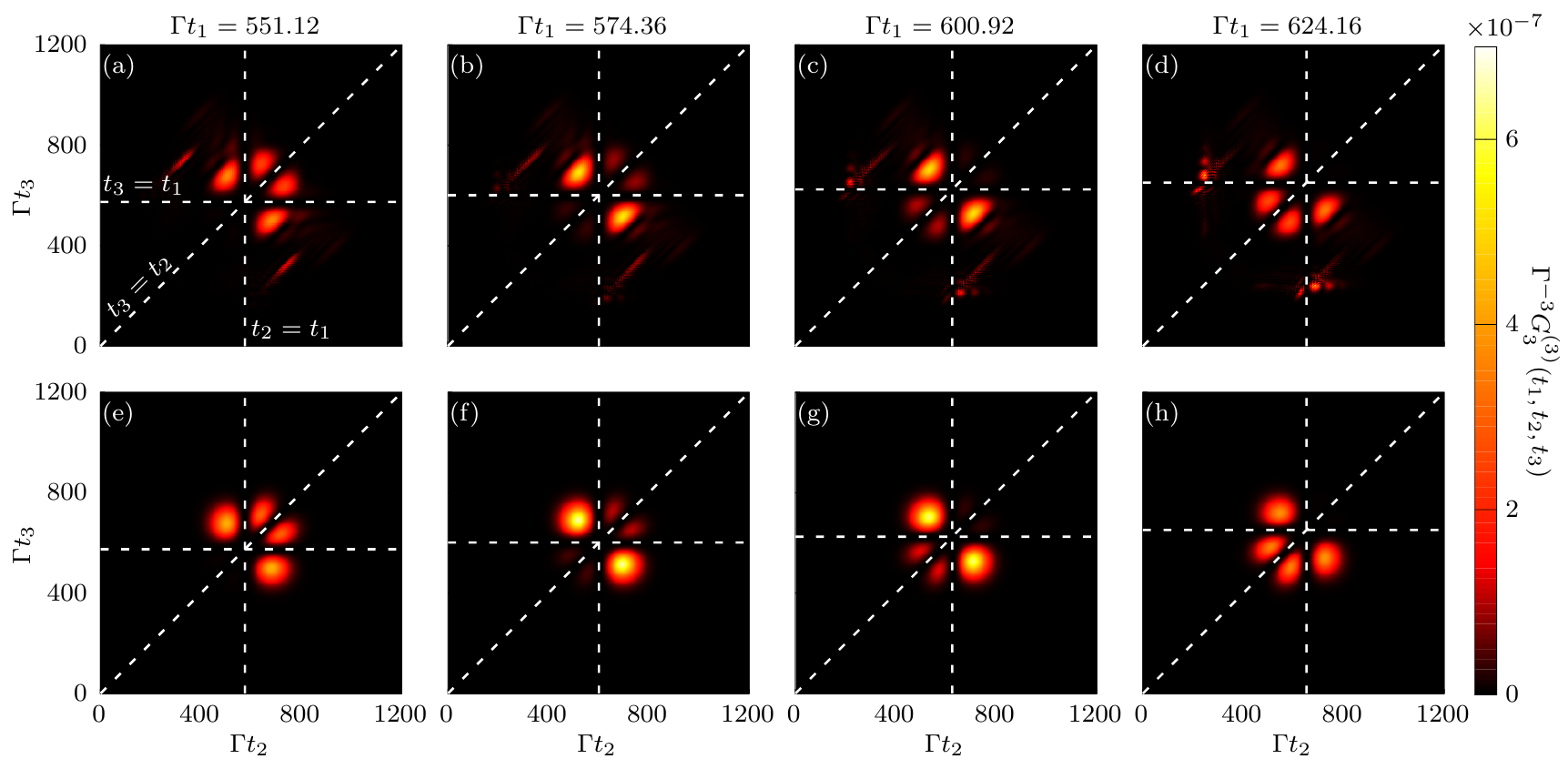}
	\caption{Three-photon density $G_3^{(3)}(t_1,t_2,t_3)$ of an incident three-photon pulse for $N = 60$ are shown in panels (a)--(d), where $t_1$ is fixed at different values indicated by the vertical and horizontal dashed lines. The pulse width and system parameters are the same as in Fig. \ref{fig:3photon-G2}(a). Corresponding results for a reference pulse of the type \eqref{eq:E-math} with the same parameters as in Fig. \ref{fig:3photon-G2}(b) are shown in panels (e)--(h). All dashed lines indicate equal times, and the three-photon density is suppressed at these times.}
	\label{fig:3photon-G3}
\end{center}
\end{figure*} 

\section{Self-Ordering of multiple photons} \label{sec:multi-pulse}
Having established the emergence of a strong effective repulsion, capable of fragmenting an initial two-photon product state into a correlated train of individual photons, we are now in a position to analyze the onset of this process for more than two photons. 
Fig. \ref{fig:3photon-G2}(a) shows the second-order photon density $G_3^{(2)}(t_1,t_2)$ of the transmitted light from a long three-photon pulse with $\Gamma\sigma = 80$. Importantly, one still finds strong antibunching, indicating that photon repulsion  remains effective also for three simultaneously interacting photons. Furthermore, the obtained two-photon density displays the notable additional features of a triple peak structure in each of the two main maxima in Fig. \ref{fig:3photon-G2}(a), which indicates self-ordering of the photons into a regular train of three photons. For a more direct illustration of this conclusion, we consider the following ansatz
\begin{widetext}
	\begin{equation}
		\begin{split}
			\mathcal{E}_3(t_1,t_2,t_3) &= C\exp\left[-\frac{(t_c-\mu_c)^2}{2\sigma_c^2}\right]\frac{|t_1-t_2|^p|t_2-t_3|^p|t_3-t_1|^p}{\sigma_{r1}^{2p}\sigma_{r2}^p} \\	
				&\quad \times\bigg\{\exp\left[-\frac{(t_1-t_2)^2}{2\sigma_{r1}^2} - \frac{(t_2-t_3)^2}{2\sigma_{r1}^2} - \frac{(t_3-t_1)^2}{2\sigma_{r2}^2}\right] + (t_1\leftrightarrow t_2) + (t_2\leftrightarrow t_3)\bigg\}.
		\end{split}
		\label{eq:E-math}
	\end{equation}
\end{widetext}
as a correlated three-photon reference amplitude. It describes a three-photon train of regularly spaced photons, and the explicit form has similarities to the Laughlin ansatz for the wave function of electrons in a magnetic field \cite{laughlin83}. Here $t_c \equiv (t_1+t_2+t_3)/3$, $\mu_c$ defines the center of the pulse, the overall pulse duration is determined by $\sigma_c$, while $\sigma_{r1}$ and $\sigma_{r2}$ define the distance between the photons. Choosing $\sigma_{r2} = 2\sigma_{r1}$ then yields a three-photon train of regularly spaced photons. As demonstrated in Fig. \ref{fig:3photon-G2}, this form of a correlated three-photon train indeed closely resembles the generated photon state as it propagates through the emitter chain.

One can verify this result more directly by considering the three-photon density $G_3^{(3)}(t_1,t_2,t_3)$, which is shown in Figs. \ref{fig:3photon-G3}(a)--(d) for consecutive time slices corresponding to different values of the detection time $t_1$ of one of the three photons.  Once more, the depicted three-photon density confirms the highly antibunched nature of the transmitted multiphoton state, as indicated by the low density along the dashed lines in Fig. \ref{fig:3photon-G3}. The numerically obtained three-photon density agrees remarkably well with our ansatz Eq. \eqref{eq:E-math} for a wave function of a regular three-photon train [Figs. \ref{fig:3photon-G3}(e)--(h)], and hence the chain facilitates three-photon self-ordering to a regular train.

\begin{figure}[b!]
\begin{center}
	\includegraphics[width=\columnwidth]{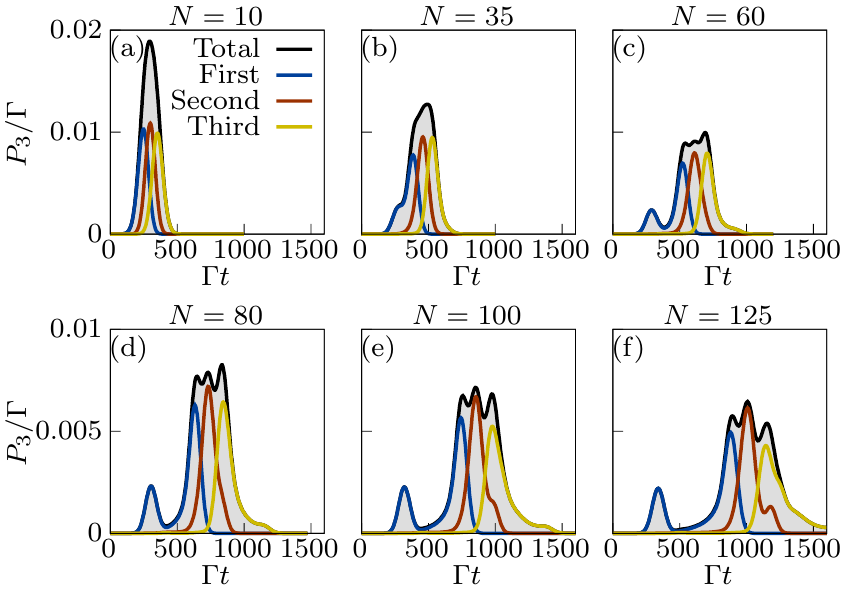}
	\caption{The onset of three-photon fragmentation is indicated by the evolution of total power (black lines and shaded area) and the power of the first (blue lines), second (red lines), and third (yellow lines) photons for an outgoing three-photon pulse with increasing chain length of (a) $N = 10$, (b) $N = 35$, (c) $N = 60$, (d) $N = 80$, (e) $N = 100$, and (f) $N = 125$ emitters. The results are obtained for the same pulse width and system parameters as Figs. \ref{fig:3photon-G2}(a) and \ref{fig:3photon-G3}(a)--(d).}
	\label{fig:power-3photon}
\end{center}
\end{figure}

Expanding on the analysis of the two-photon dynamics in Section \ref{sec:two-pulse}, we study the dynamics of the single-photon densities
\begin{align}
	P_{3,1} &= \int_t^{\infty} dt_2 \int_t^{\infty} dt_3\, G_3^{(3)}(t,t_2,t_3),  \label{eq:P3,first}\\
	P_{3,2} &= 2\int_{-\infty}^t dt_2\int_t^{\infty} dt_3\, G_3^{(3)}(t,t_2,t_3), \label{eq:P3,second}\\
	P_{3,3} &= \int_{-\infty}^t dt_2\int_{-\infty}^t dt_3\, G_3^{(3)}(t,t_2,t_3), \label{eq:P3,third}
\end{align}
which, analogously to Eqs. \eqref{eq:Pfirst} and \eqref{eq:Psecond}, describe the power profiles of the first, second, and third photon in the propagating three-photon pulse, respectively. Although the overall dynamics requires a larger number of emitters compared to the two-photon pulse discussed in the previous section, the transmitted photon densities shown in Fig. \ref{fig:power-3photon} clearly reveals the onset of a multiphoton fragmentation process into a train of single photons.

\begin{figure*}
\begin{center}
	\includegraphics[width=\textwidth]{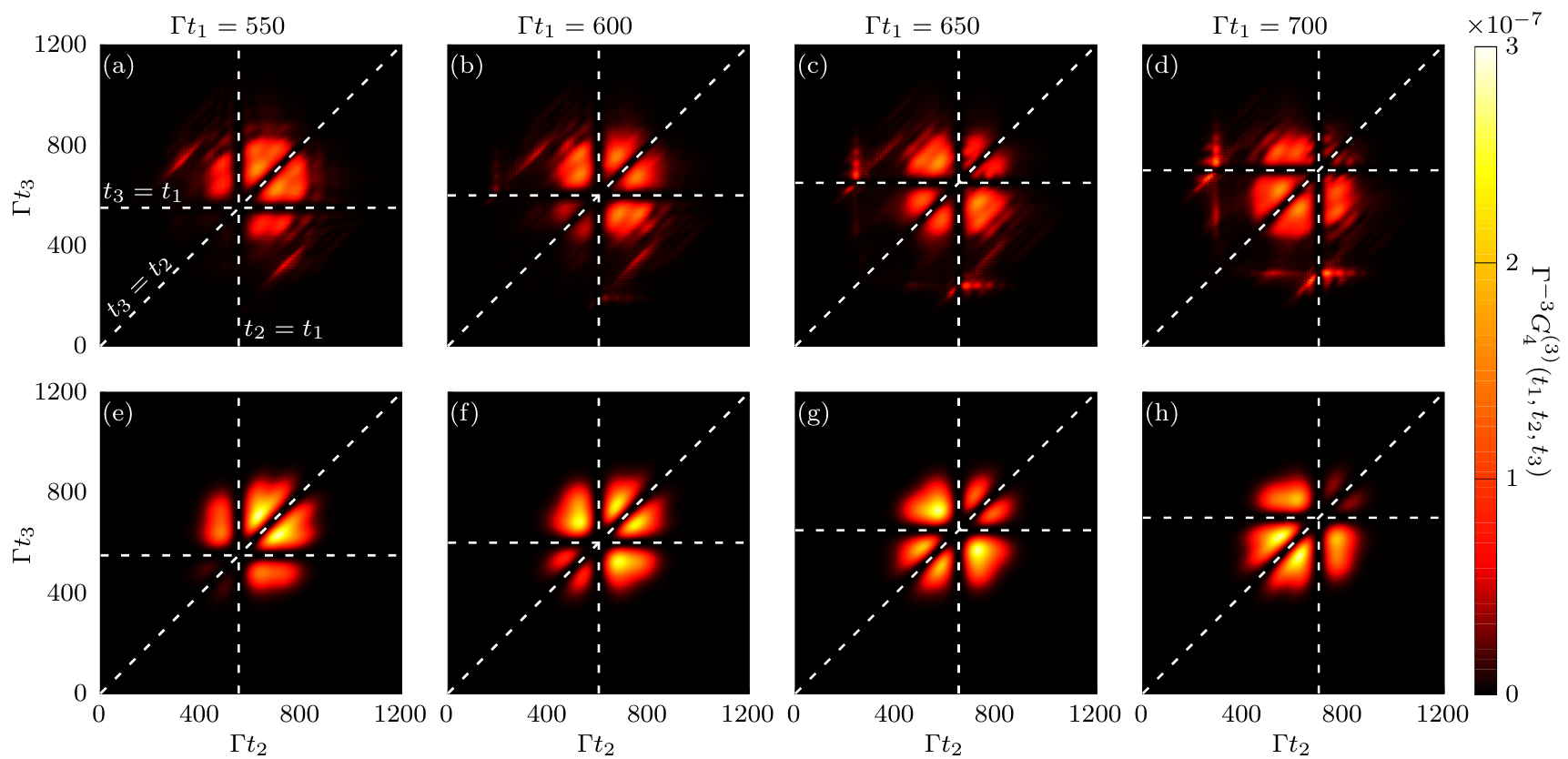}
	\caption{Three-photon density $G_4^{(3)}(t_1,t_2,t_3)$ of a four-photon output of a chain with $N = 60$ emitters [panels (a)--(d)], where the incident pulse width is $\Gamma\sigma = 120$, and the remaining parameters are $\Omega = 0.35\Gamma$, $\Delta = 0.5\Gamma$ and $\bar{\Delta} = 0.4\Gamma$. Corresponding results for a pulse of the type \eqref{eq:E4-math} are shown in panels (e)--(h), where $p = 1.2$, $\Gamma\mu_c = 575$, $\Gamma\sigma_c = 70$, $\Gamma\sigma_{r1} = 90/\sqrt{p}$, $\sigma_{r2} = 2\sigma_{r1}$ and $\sigma_{r3} = 3\sigma_{r1}$ have been applied. The white dashed lines indicate where two times are equal.}
	\label{fig:4photon-G3}
\end{center}
\end{figure*}

While an extension of the described simulations towards higher photon numbers becomes exceedingly numerically demanding, we complete the discussion with a brief analysis of numerical results for four incident photons and a chain of $N=60$ emitters. A comparison of the calculated three-photon density $G_4^{(3)}$ with the corresponding regular-train ansatz for a four-photon state (see Appendix \ref{sec:4photonAnsatz}) is shown in Fig.~\ref{fig:4photon-G3}. The calculations demonstrate that the strong antibunching persists among all four photons and the striking agreement between panels (a)--(d) and (e)--(h) illustrates the spontaneous multiphoton self-ordering dynamics into regular trains of individual photons.

\section{Experimental Implementation} \label{sec:superatoms}
Experiments with cold atoms \cite{mitsch14,petersen14} and quantum dots \cite{sollner15,Coles16,JalaliMehrabad20,dewhurst10} have demonstrated chiral coupling to nanophotonic waveguides \cite{chiral-qed-review}, and multilevel driving can also be implemented in both systems \cite{warburton13,LeKien15}. While quantum dot experiments can reach near-perfect photon coupling to a single waveguide mode \cite{arcari14,sollner15}, realizing arrays of multiple identical emitters has been naturally difficult with solid-state components \cite{Kim18,Grim19}. While this is readily possible with atomic systems \cite{mitsch14,petersen14,prasad20}, reaching high coupling efficiencies remains challenging, but is possible with optical resonators \cite{hallett21} or by operating close to a bandgap of photonic crystal waveguides \cite{goban14,Goban15,hood16}.

Alternatively, strong unidirectional coupling to a single photonic mode can be realized with mesoscopic atomic ensembles \cite{Saffman02,paris-mandoki17}. Hereby, the coupling to the single transverse mode of an incident field is collectively enhanced as photons interact with a large number of atoms in the ensemble. While the simultaneous interaction with many atoms diminishes the intrinsic nonlinearity of single atoms, the strong saturation can be restored by exciting atoms to high-lying Rydberg states \cite{pritchard10,sevincli11,Petrosyan11,gorshkov11,dudin12,Peyronel12}. At principal quantum numbers of $n\gtrsim100$, Rydberg atoms possess very strong van der Waals interactions that exceed interactions between ground state atoms by many orders of magnitude \cite{Gallagher08}. The associated level shift induced by the atomic interactions can be sufficiently large to block all but a single excitation in the ensemble \cite{Lukin01}, which then behaves as an effective single two-level emitter in the form of a so-called superatom whose transition is saturated at the single-photon level. Numerous experiments have demonstrated and exploited this Rydberg blockade effect for different applications \cite{Saffman10,Browaeys20,murray2016-review,firstenberg2016-review}. 

\begin{figure}
\begin{center}
	\includegraphics[width=\columnwidth]{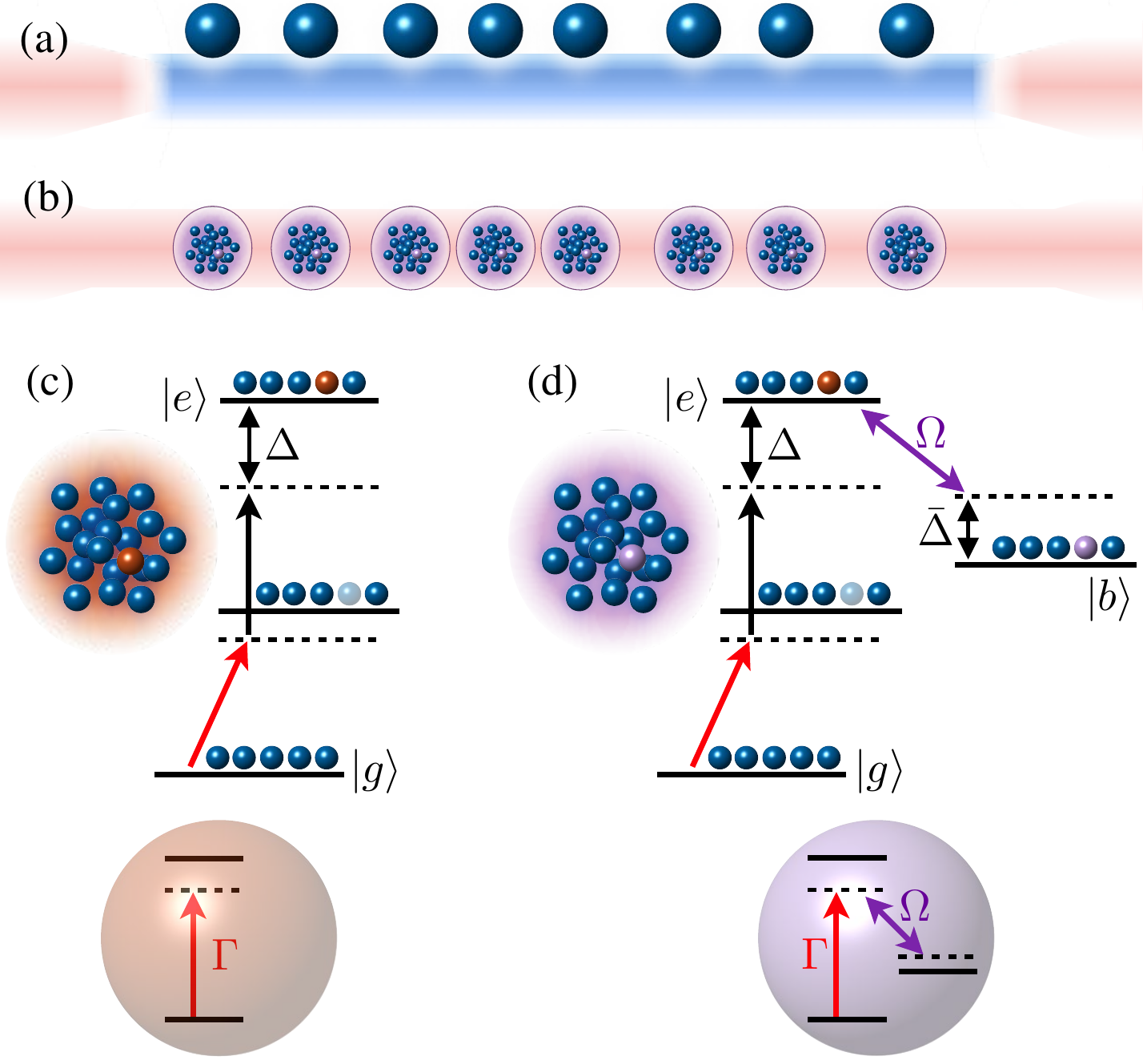}
	\caption{(a) A chiral chain of waveguide-coupled emitters can be (b) effectively realized with a one-dimensional array of mesoscopic atomic ensembles, involving high-lying Rydberg states.  As illustrated in panel (c), incident photons can be coupled to a Rydberg-state transition via off-resonant two-photon excitation  \cite{paris-mandoki17}, while the strong van der Waals interaction between Rydberg atoms prevents the generation of multiple excitations in each cloud. Due to this Rydberg blockade effect the mesoscopic ensemble effectively behaves like a single saturable two-level emitter. (d) By coupling the laser-excited Ryberg state $|e\rangle$ to another Rydberg state $|b\rangle$ via an external microwave field, it is possible to create a chiral chain of effective three-level emitters as we have studied in this work.} \label{fig:superatom}
\end{center}
\end{figure}

Reference \cite{paris-mandoki17} demonstrated strong coupling between a single photonic mode and a Rydberg super-atom containing $N\sim10^4$ cold rubidium atoms using an off-resonant two-photon transition [see Fig. \ref{fig:superatom}(c)], and direct single-photon coupling to Rydberg states is also possible \cite{Zeiher16,Zeiher17,Hollerith19}. Here, the collective enhancement of the emission rate $\Gamma\propto N$ into the incident optical mode compared to the photon-loss rate $\gamma$ makes it possible to achieve large coupling efficiencies $\beta=\Gamma/(\Gamma+\gamma)$. Experiments suggest $\beta>0.8$ \cite{paris-mandoki17}, while currently observed coupling efficiencies are confined to lower values by external noise, residual thermal motion of the atoms, and dipolar excitation exchange in the disordered ensemble \cite{kumlin18,stiesdal20,stiesdal21}, which give rise to dephasing that can be reduced at lower temperatures. The effects of atomic motion and disorder can both be eliminated with ordered lattices of single atoms \cite{zhang2021,morenocardoner2021}, and coupling efficiencies $\beta\rightarrow1$ appear feasible in future experiments with mesoscopic ensembles or ordered atomic lattices.

Photon coupling to multiple such superatoms is possible with arrays of dipole traps \cite{stiesdal21} and effectively realizes a waveguide QED setup in which a single spatial photonic mode is coupled to a chain of two-level emitters with high efficiency, as illustrated in Figs. \ref{fig:superatom}(a) and (b). Here, we propose an extension of this approach to realize effective three-level emitters in the form of three-level superatoms. The idea is illustrated in Fig. \ref{fig:superatom}(d), whereby a microwave field is used to couple the optically excited Rydberg state to another Rydberg state. Owing to the typically large transition dipole moment of this Rydberg-Rydberg state transition, the corresponding Rabi frequency $\Omega$ is broadly tunable by varying the microwave intensity and can cover the parameter range considered in this work. Importantly, both involved Rydberg states feature strong interactions, which retain the excitation blockade and still prohibit the excitation of more than a single atom within an ensemble. Hence, this setting realizes mesoscopic three-level super-atoms with a strong and well controllable coupling to a single photonic mode. 

\section{Summary and conclusions} \label{sec:conclusion}
In conclusion, we have investigated the propagation dynamics of multiple photons through chiral chains of three-level emitters, and demonstrated the emergence of a strong effective photon-photon repulsion in this system. The repulsive nonlinearity arises from the saturation of the individual emitters, which is remarkable since such saturation effects typically lead to a nonlinear focussing of light in two-level optical media. Here, we have found that the additional optical driving can eliminate the bound state in the two-photon scattering off a single three-level emitter and generate a repulsive interaction between individual photons. Indeed, the resulting interactions are strong enough to convert incident single-mode product states into strongly correlated states of light, inducing a fragmentation into single-photon pulses or the formation of highly ordered multiphoton states.

As repulsive nonlinearities are generally scarce and difficult to realize, the presented results suggest an appealing platform to explore optical self-organization at the quantum level of individual photons. The found dynamics of few-photon pulses motivates future work on efficient numerical schemes or approximate treatments to further study the emergence of order and potential crystallization of light in systems of long chains and many photons. Considering the profound effects of the additional control field, it will be interesting to investigate multilevel driving schemes such as four-level emitters in $N$-type configurations, which can generate photon interactions under conditions of electromagnetically induced transparency \cite{zheng11,zheng12} and may thus yield a different type of photon interaction. 

We have discussed different platforms to experimentally study the nonlinear photon transport explored in this work. Rydberg-atom ensembles offer a promising realization of unidirectional quantum emitters with high single-mode coupling efficiencies $\beta = \Gamma/(\Gamma+\gamma)$, which may be improved further in future experiments. In order to study experimental prospects, it will be important in future theoretical work to explore the impact of potential imperfections such as nonunity emitter-waveguide coupling, imperfect chirality, and dephasing.
 
Although we have focused here on fundamental aspects of the photon dynamics, the demonstrated interaction mechanism can also be interesting for quantum optics applications. For example, future work to further increase the efficiency of the demonstrated photon-pulse fragmentation by optimizing parameters and shaping the incident photon pulse may permit to explore the generation and dynamical emergence of single-photon pulse trains from classical light.

\begin{acknowledgments}
We thank Lida Zhang, Jan Kumlin, Klaus M\o{}lmer and Anders S\o{}rensen for helpful comments and fruitful discussions. 
This work was supported by the Carlsberg Foundation through the 'Semper Ardens' Research Project QCooL, by the DFG through the SPP1929, by the European Commission through the H2020-FETOPEN project ErBeStA (No. 800942), and by the DNRF through a Niels Bohr
Professorship to T.P. and the Center of Excellence “CCQ” (Grant agreement no.: DNRF156).\end{acknowledgments}

\appendix

\section{Details for the Analysis of the Two-Photon Fragmentation} \label{sec:delay}
In this appendix, we provide further details for the analysis of the fragmentation of an incident uncorrelated two-photon pulse described in Section \ref{sec:two-pulse}. Specifically, we provide the spectral analysis of the two-photon output and show how we obtain the solid lines in Fig. \ref{fig:power-2photon}, which supports the picture of a finite range repulsive photon-interaction generating a relative momentum kick.

\begin{figure}[b!]
\begin{center}
	\includegraphics[width=\columnwidth]{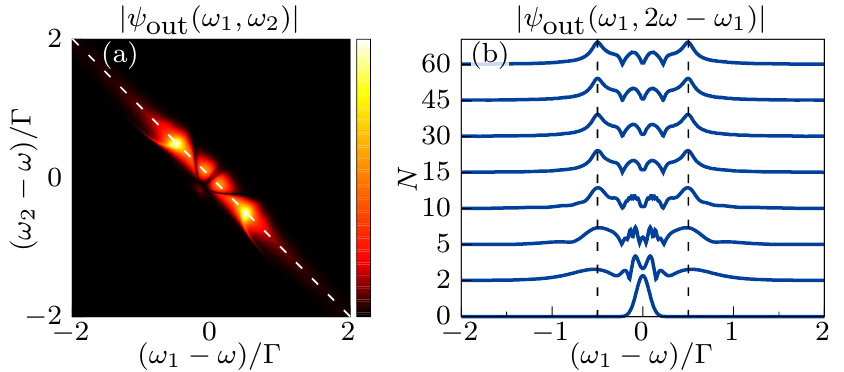}
	\caption{(a) Fourier transform $\psi_{\rm out}(\omega_1,\omega_2)$ of the outgoing two-photon amplitude for $N = 30$. The main contribution is located along the energy conserving line, indicated by the white dashed line, where the sum of the frequencies $\omega_1$ and $\omega_2$ equals twice the incident carrier frequency $\omega$. (b) Fourier transform for various numbers of emitters $N$ along the energy conserving line in panel (a). After scattering off 5--10 emitters peaks emerge around $\omega_1 = \omega \pm 0.5\Gamma$, indicated by the black dashed lines, and these peaks persists as the chain length is increased. The results are obtained for $\Omega = 0.25\Gamma$, $\Delta = 0.5\Gamma$, and $\bar{\Delta} = 0.2\Gamma$.}
	\label{fig:fourier}
\end{center}
\end{figure}

For the spectral analysis of the output, we compute the fast Fourier transform of the two-photon amplitude to determine the frequencies of the photons after scattering. To this end, the two-photon density is insufficient, but we must retain phase information of the transmitted two-photon amplitude. As required by energy conservation, we find that the frequencies mainly contributing to the Fourier transform obeys $\omega_1 + \omega_2 \approx 2\omega$, as illustrated in Fig. \ref{fig:fourier}(a), where $\omega_1$ and $\omega_2$ are the frequencies of the outgoing photons and $\omega$ is the carrier frequency of the incident photons. By studying the Fourier transform at frequencies exactly obeying the energy conservation for various chain lengths [Fig. \ref{fig:fourier}(b)], it becomes evident that two peaks emerge at frequencies shifted by around $\pm 0.5\Gamma$ relative to the incident frequency. Hence, the photon interaction in the initial part of the chain generates frequency shifts of the photons, which afterwards remains constant upon scattering off additional emitters. 

Based on these frequency shifts, it is natural to assert that the further increase of distance between the photons, seen in Fig. \ref{fig:power-2photon}, when scattering off additional emitters, originates from different single-photon group velocities for the two different frequencies. To verify this assertion, we compare our results to an analytical expression for the single-photon group velocity for propagation through the chain. This expression provides the slope of the lines in Fig. \ref{fig:power-2photon}, which allows comparison to the two-photon results by ensuring the correct departure times from the chain for $N = 10$ emitters.

To obtain the group velocity, we find the single-photon delay imposed by each emitter. To this end, we note that an incident single-photon state in the rotating frame is given by
\begin{equation}
	\ket{\psi_{\rm in}} = \int dx\, \psi_1(x) \hat{a}^{\dagger}(x)\ket{0}, \label{eq:psi-in}
\end{equation}
where $\psi_1(x)$ is the single-photon wave function and $\ket{0}$ is the vacuum state. Changing basis to the energy (or, equivalently, momentum) eigenstates $\ket{\delta} = (2\pi)^{-1/2}\int dx\, e^{i\delta x}\hat{a}^{\dagger}(x)\ket{0}$, where $\delta = \tilde{\omega} - \omega$ is the detuning from the carrier frequency, yields
\begin{equation}
	\ket{\psi_{\rm in}} = \int d\delta\, A(\delta)\ket{\delta},
\end{equation}
where the connection between $A$ and $\psi_1$ is provided by the Fourier transform
\begin{equation}
	\psi_1(x) = \frac{1}{\sqrt{2\pi}}\int d\delta\, A(\delta)e^{i\delta x}.
\end{equation}
Upon scattering off a single emitter, the energy eigenstates are multiplied by the transmission coefficient \cite{witthaut10}
\begin{equation}
	T_{\delta} = \frac{\Omega^2 - (\Delta-\delta)(\bar{\Delta}-\delta) - i\Gamma(\bar{\Delta}-\delta)/2}{\Omega^2 - (\Delta-\delta)(\bar{\Delta}-\delta) + i\Gamma(\bar{\Delta}-\delta)/2}. \label{eq:T}
\end{equation}
As there is no dissipation, we can write $T_{\delta} = e^{i\theta_{\delta}}$, where
\begin{equation}
	\theta_{\delta} = -2\arctan\left[\frac{\Gamma(\bar{\Delta}-\delta)}{2\left(\Omega^2 - (\Delta-\delta)(\bar{\Delta}-\delta)\right)}\right].
\end{equation}
A Taylor expansion to first order in $\delta$ yields
\begin{equation}
	\theta_{\delta} \simeq \theta_0 + \tau_d \delta, \label{eq:theta-first-order}
\end{equation}
where $\theta_0 = -2\arctan[\Gamma\bar{\Delta}/(2(\Omega^2-\Delta\bar{\Delta}))]$ and
\begin{equation}
	\tau_d = \Gamma\frac{\Omega^2 +\bar{\Delta}^2}{(\Omega^2 - \Delta\bar{\Delta})^2 + \Gamma^2 \bar{\Delta}^2/4}.
\end{equation}
The output after scattering off one emitter is now found to first order as
\begin{align}
	\ket{\psi_{\rm out}} &= \int d\delta\, T_{\delta} A(\delta) \ket{\delta} \nonumber\\
		 &= e^{i\theta_0} \int dx\, \frac{1}{\sqrt{2\pi}} \int d\delta\, A(\delta) e^{i\delta x}e^{i\tau_1 \delta} \hat{a}^{\dagger}(x)\ket{0} \nonumber\\
		 &= T_0 \int dx\, \psi_1(x+\tau_1)\hat{a}^{\dagger}(x)\ket{0}, \label{eq:psi-out}
\end{align}
where $T_0 = e^{i\theta_0}$ is the transmission coefficient at the carrier frequency. By comparing Eqs. \eqref{eq:psi-in} and \eqref{eq:psi-out}, it is evident that, independent of the incident pulse shape, the outgoing pulse after scattering off one emitter has acquired a phase given by the transmission coefficient at the carrier frequency, and it has been shifted in the negative $x$-direction, i.e. delayed, by $\tau_d$. In the situation with $N$ emitters, each emitter causes a delay of $\tau_d$ resulting in a total delay of $N\tau_d$. We therefore define a group velocity as $v_g = \tau_d^{-1}$, which yields Eq. \eqref{eq:vg} and provides the number of emitters the photon pass per unit time. Hereby, $v_g$ is the slope of the lines in Fig. \ref{fig:power-2photon}, while the intercepts are chosen to ensure agreement with the simulated departure times for $N = 10$ emitters, where the photons have separated in Fig. \ref{fig:power-2photon}.

\section{Four-Photon Pulse Train}\label{sec:4photonAnsatz}
Here we give the explicit form of the wave function ansatz to describe the regular train of four photons in Figs.~\ref{fig:4photon-G3}(e)--(h). Analogous to the analysis of the three-photon pulse in Eq. \eqref{eq:E-math}, we use the following form for the four-photon amplitude
\begin{widetext}
	\begin{equation}
		\begin{split}
			\mathcal{E}_4(t_1,t_2,t_3,t_4) &= C\exp\left[-\frac{(t_c-\mu_c)^2}{2\sigma_c^2}\right]\frac{|t_{12}|^p|t_{23}|^p|t_{34}|^p|t_{13}|^p |t_{24}|^p |t_{14}|^p}{\sigma_{r1}^{6p}\sigma_{r2}^{2p}\sigma_{r3}^p} \\	
				&\quad \times\bigg\{\exp\left[-\frac{t_{12}^2 + t_{23}^2 + t_{34}^2}{2\sigma_{r1}^2} - \frac{t_{13}^2 + t_{24}^2}{2\sigma_{r2}^2} - \frac{t_{14}^2}{2\sigma_{r3}^2}\right] + \textup{permutations}\bigg\},
		\end{split}
		\label{eq:E4-math}
	\end{equation}
\end{widetext}
where $t_c = (t_1+t_2+t_3+t_4)/4$, $t_{ij} = t_i - t_j$, and ``permutations'' include all necessary permutations of $t_1$, $t_2$, $t_3$, and $t_4$ to adhere to the requirements of bosonic symmetry. In analogy to the three-photon case, we use $\sigma_{r3} = 3\sigma_{r2}/2 = 3\sigma_{r1}$, which corresponds to a state with four equidistant photons. For the comparison to our numerical simulations in Fig.~\ref{fig:4photon-G3}, we use $p = 1.2$, $\Gamma\mu_c = 575$, $\Gamma\sigma_c = 70$, and $\Gamma\sigma_{r1} = 90/\sqrt{p}$.

\end{document}